\documentclass{elsart} 

\usepackage[square,comma,numbers]{natbib}
\usepackage{graphicx}
\usepackage{pxfonts}
\usepackage{lineno}

\usepackage{amssymb}
\usepackage{footnote}
\usepackage{multirow}
\usepackage{varwidth}
\usepackage{array}
\usepackage{epstopdf}

\usepackage{grffile} 
\usepackage{makecell} 
\usepackage{lscape}

\journal{}

\begin{document}

\thispagestyle{empty}
\begin{Large}
	\textbf{DEUTSCHES ELEKTRONEN-SYNCHROTRON}
	
	\textbf{\large{Ein Forschungszentrum der
			Helmholtz-Gemeinschaft}\\}
\end{Large}

DESY 16-167

August 2016

\begin{eqnarray}
\nonumber &&\cr \nonumber && \cr \nonumber &&\cr
\end{eqnarray}
\begin{eqnarray}
\nonumber
\end{eqnarray}
\begin{center}
	\begin{Large}
		\textbf{Circular polarization opportunities at the SASE3 undulator line of the European XFEL}
	\end{Large}
	\begin{eqnarray}
	\nonumber &&\cr \nonumber && \cr
	\end{eqnarray}
	
	\begin{large}
		Svitozar Serkez, Gianluca Geloni,
	\end{large}
	\textsl{\\European XFEL GmbH, Hamburg}
	\begin{large}
		
		Vitali Kocharyan and Evgeni Saldin
	\end{large}
	\textsl{\\Deutsches Elektronen-Synchrotron DESY, Hamburg}
	\begin{eqnarray}
	\nonumber
	\end{eqnarray}
	\begin{eqnarray}
	\nonumber
	\end{eqnarray}
	ISSN 0418-9833
	\begin{eqnarray}
	\nonumber
	\end{eqnarray}
	\begin{large}
		\textbf{NOTKESTRASSE 85 - 22607 HAMBURG}
	\end{large}
\end{center}
\clearpage
\newpage

\begin{frontmatter}



\title{Circular polarization opportunities at the SASE3 undulator line of the European XFEL}


\author[XFEL]{Svitozar Serkez,}
\author[XFEL]{Gianluca Geloni,}
\author[DESY]{Vitali Kocharyan,}
\author[DESY]{Evgeni Saldin}
\address[XFEL]{European XFEL GmbH, Hamburg, Germany}
\address[DESY]{Deutsches Elektronen-Synchrotron (DESY), Hamburg, Germany}

\begin{abstract}
XFELs provide X-ray pulses with unprecedented peak brightness and ultrashort duration. They are usually driven by planar undulators, meaning that the output radiation is linearly polarized. For many experimental applications, however, polarization control is critical: besides the ability to produce linearly polarized radiation, one often needs the possibility of generating circularly polarized radiation with a high, stable degree of polarization. This may be achieved by using a first part of the XFEL undulator to produce bunching and then, by propagating the the bunched beam through an ``afterburner'' -- a short undulator with tunable polarization, where only limited gain takes place. One of the issues that one needs to consider in this case is the separation of the circularly polarized radiation obtained in the radiator from the linearly polarized background produced in the first part of the FEL. In this article we review several methods to do so, including the inverse tapering technique. In particular, we use the Genesis FEL code to simulate a case study pertaining to the SASE3 FEL line at the European XFEL with up-to-date parameters and we confirm that a high degree of circular polarization is expected. Moreover, we propose to further improve the effectiveness of the inverse tapering technique either via angular separation of the linearly polarized radiation or strongly defocusing it at the sample position. In this way we exploit the unique flexibility of the European XFEL from both the electron beam and the photon beam optics side.
\end{abstract}

%
%

%
\end{frontmatter}


\section{ Introduction }

Controlling the polarization of X-ray FEL pulses is critical for many experiments. In particular, a number of FEL applications in the soft X-ray range of the electromagnetic spectrum require the possibility of switching between left- and right-handed circularly polarized pulses with high, stable degree of polarization.

However, presently, most XFELs are based on planar undulators, meaning that the output radiation is mainly linearly polarized in the direction orthogonal to the electron acceleration, with  a very small power fraction in the other direction~\cite{Geloni2015c}. A straightforward solution to obtain full polarization control is to rely on APPLE-like undulators~\cite{Sasaki1994,Hwang1999}, as was done for example at FERMI~\cite{Allaria2012}. This solution is nevertheless not convenient in the case of facilities already built or under construction, and is also more expensive.

A way around this issue is to use planar undulators only for inducing bunching in the electron beam. The bunched beam can then be sent through a short ``afterburner'' undulator with polarization-control capabilities. In this case, the afterburner acts as a coherent radiator emitting, to fix the ideas, circularly polarized light, while the linearly polarized radiation emitted during the bunching process is only a detrimental byproduct, and should be separated from the main circularly polarized pulse, or suppressed before reaching the sample. It is natural to start the afterburner before the bunching saturates, so that the circularly polarized pulse still undergoes FEL amplification (albeit limited) in the afterburner, and will thus have higher energy than the linearly polarized one. However, even accounting for a few extra gain-lengths in the afterburner, the power ratio between the circularly polarized and the linearly polarized pulses only amounts to several units. In contrast, a ratio of about a thousand is desirable, to be sure that the purest possible degree of polarization
is achieved. Several approaches have been developed to address this issue.

One may think of exploiting the bunching at higher harmonics that develops in the electron beam near saturation. In this case, the planar undulator preceding the afterburner is tuned at a subharmonic of the target wavelength and the two pulses are separated in wavelength. However, depending on the experiment, one may still need to spectrally filter the output radiation. Moreover, this scheme cannot be used to reach the longest possible wavelengths for which the XFEL is designed. For example, if the lowest photon energy achievable is about $250$~eV, like in the case of SASE3 at the European XFEL~\cite{XTDR}, using this method one could control the polarization starting from the second harmonic only, i.e. $500$~eV.

Several other approaches to reduce the power of the linearly polarized radiation have been proposed. In~\cite{Geloni2011b} it was proposed to produce the bunching well upstream of the afterburner. Since here we discuss about soft X-rays, one usually needs only a few XFEL segments to reach the optimum bunching, and the ultrarelativistic energy in the multi-GeV range guarantees that the bunched beam can be transported without deterioration up to the afterburner. Then, due to divergence, the difference in the radiation spot sizes relevant to linearly and circularly polarized pulses can be large enough to spatially filter the linearly polarized pulse using a thin slotted foil, while the electron beam can bypass the foil through a chicane, before being dumped. In this way it was previously demonstrated that a power ratio in the order of $10^3$ can be achieved.

An elegant alternative was proposed in~\cite{Schneidmiller2013}. In that paper an asymptotic solution of the FEL equations for large negative values of the detuning parameter is used. It maximizes the electron beam bunching while minimizing the output radiation. This works, in practice, by increasing the value of the undulator parameter~$K$ while the electron beam progresses through the linear undulator -- hence the name ``inverse tapering''. At the entrance of the afterburner the beam is strongly bunched due to FEL interaction, but the linearly polarized pulse is strongly suppressed. The method was tested at the LCLS~\cite{Lutman2016}, where a contrast factor~$20$ was obtained. Similar tests at FLASH~\cite{Schneidmiller2016} demonstrated a suppression of a factor $200$, which should be also obtainable at the European XFEL. The main reason for the different performance is ascribed to the sensitivity of the inverse tapering method on the electron energy spread.


In this note we study the performance of the inverse tapering method for the SASE3 beamline of the European XFEL using the simulation code GENESIS~\cite{GENE}. In this way we can include electron beam distributions in current, emittance, energy and energy spread obtained from start-to-end beam dynamics simulations~\cite{mpy_web}, and an undulator lattice with intersections~\cite{Tschentscher2011}. We confirm that the method is capable of yielding a suppression factor in the order of a thousand. Further on, we complement our studies with several techniques based on electron and photon beam optics automatically available at SASE3, to significantly further decrease the density of the linearly polarized radiation at the sample position.

\section{FEL Simulation Results}

\begin{figure}
\centering
\includegraphics[width=0.45\linewidth]{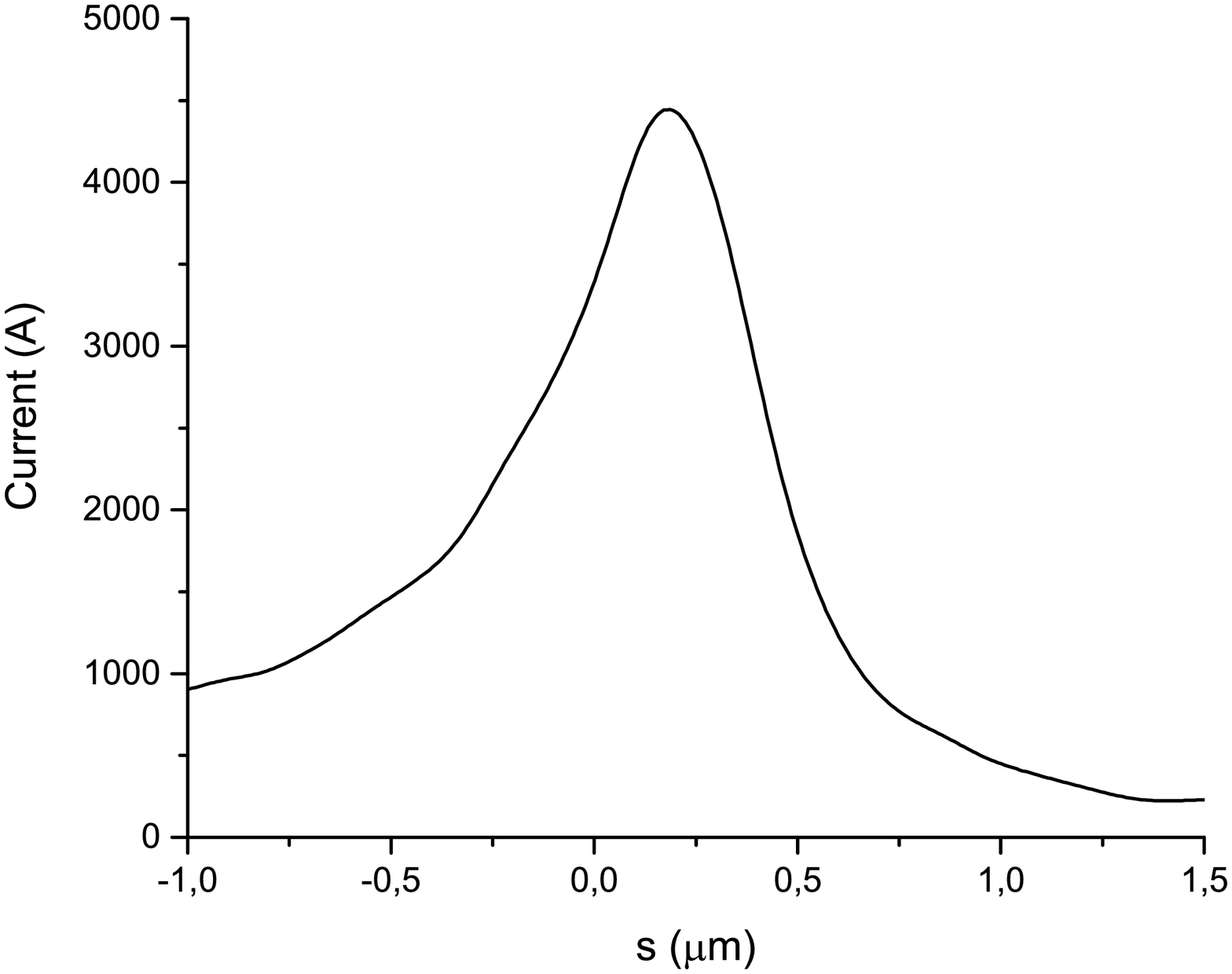}
\includegraphics[width=0.45\linewidth]{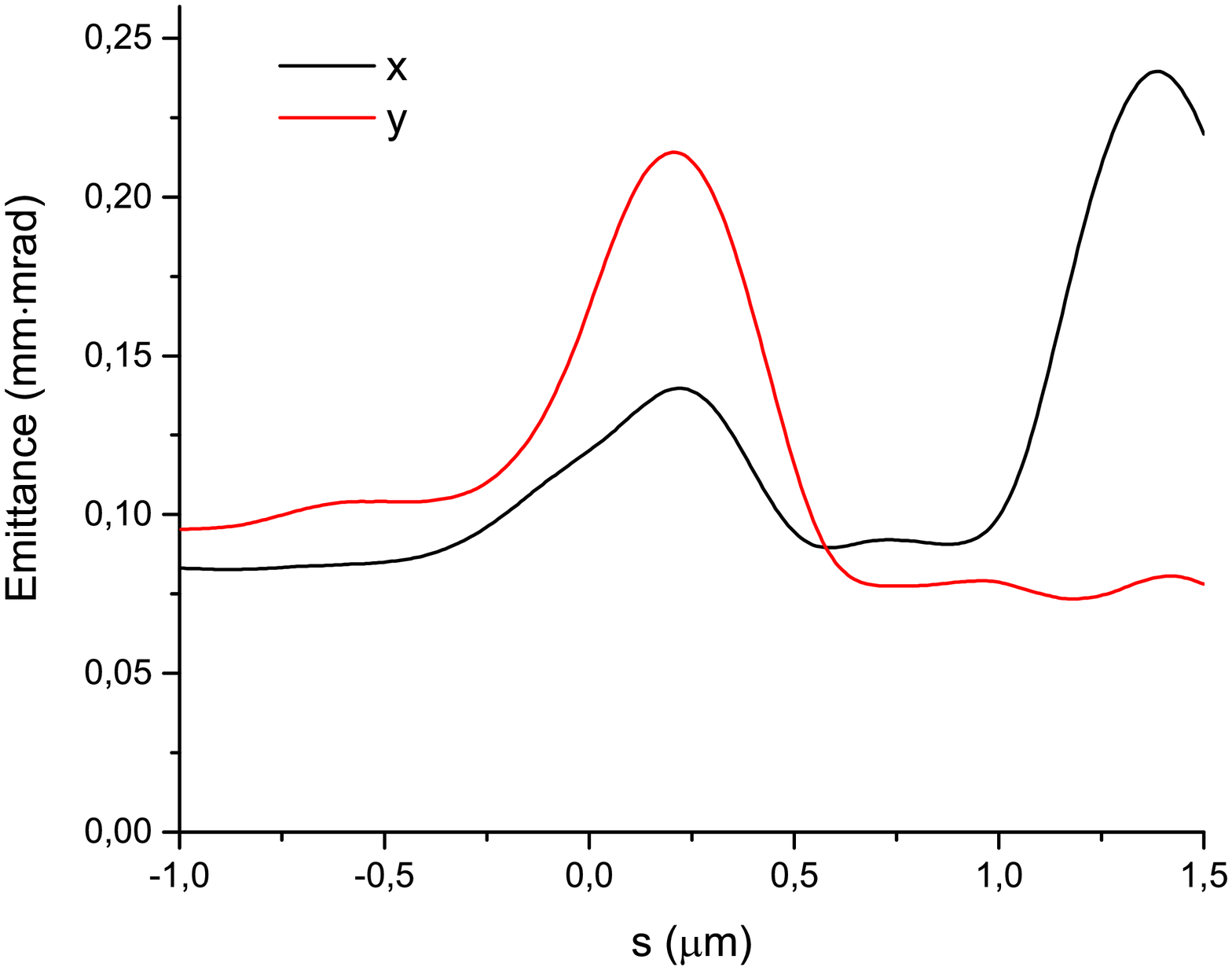}
\includegraphics[width=0.45\linewidth]{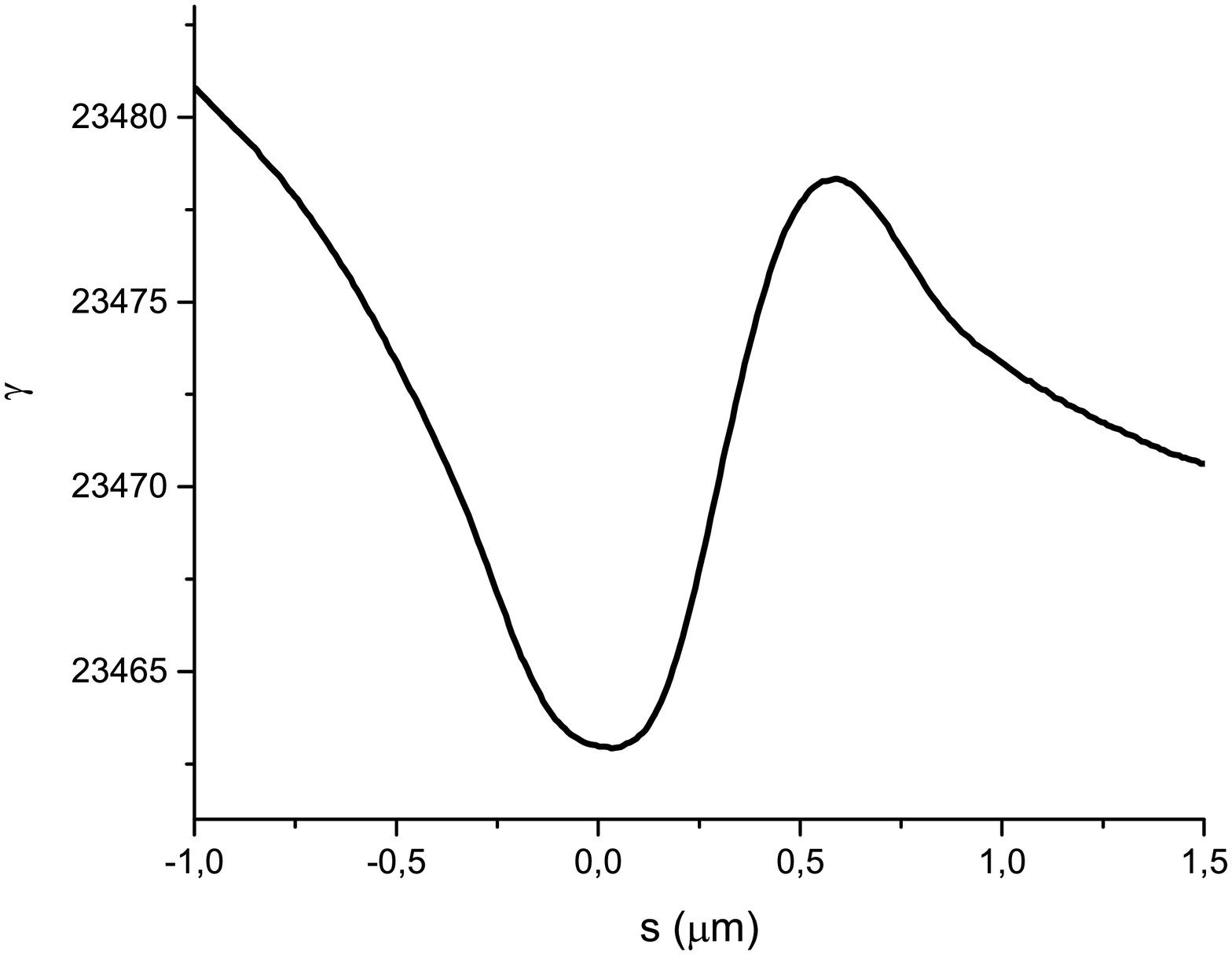}
\includegraphics[width=0.45\linewidth]{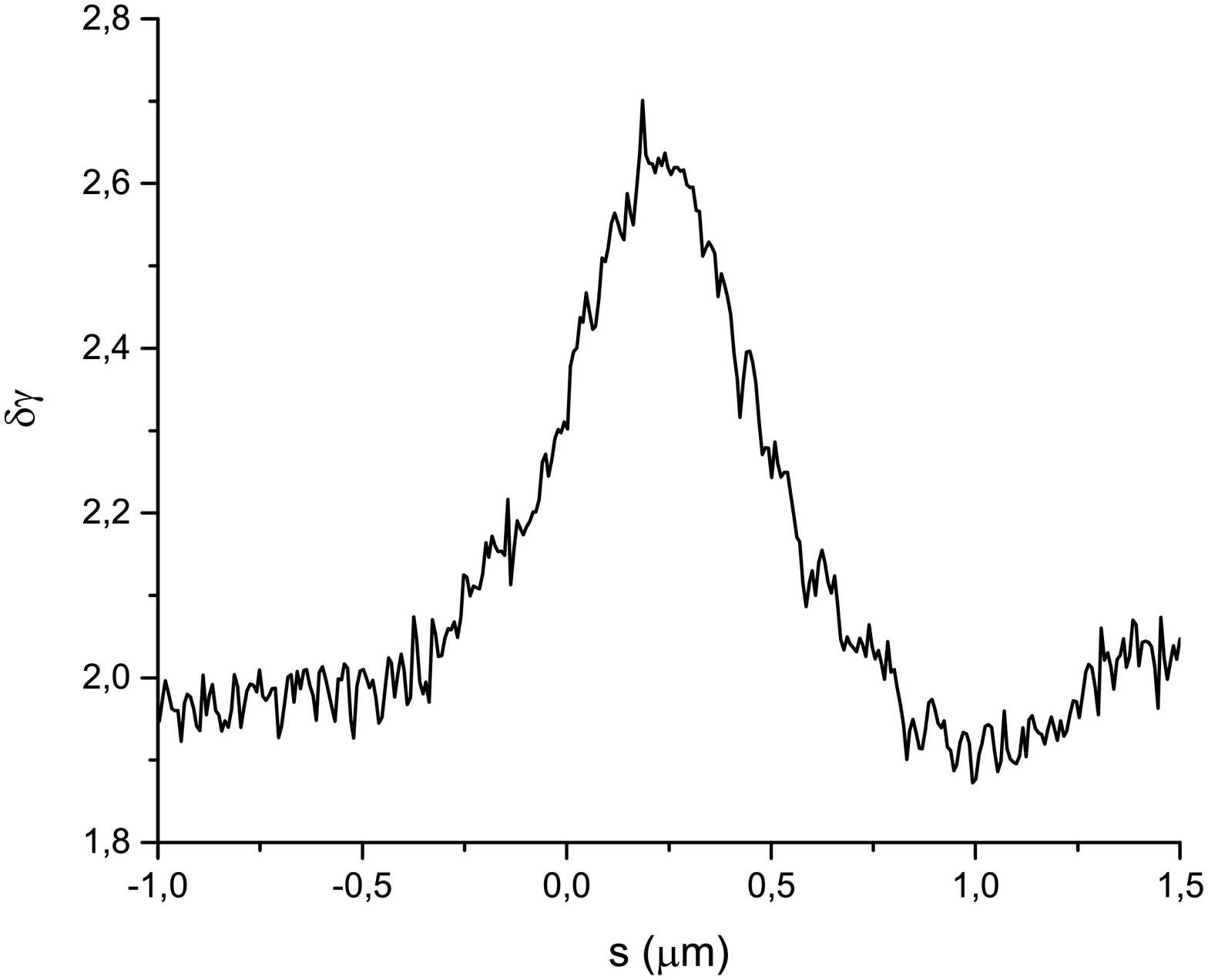}
\includegraphics[width=0.45\linewidth]{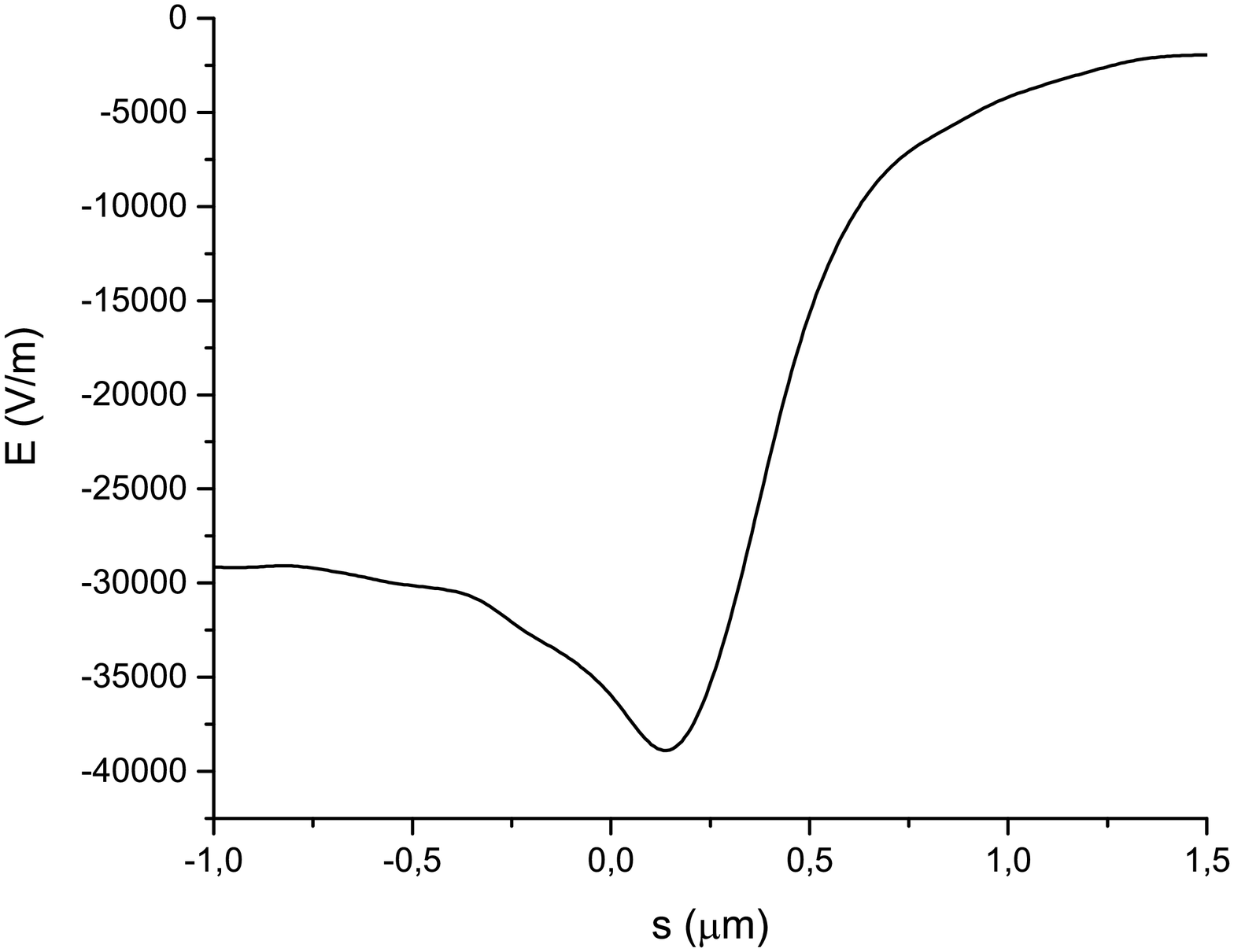}
\caption{Results from electron beam start-to-end simulations at the entrance of SASE3. (First Row, Left) Current profile. (First Row, Right) Normalized emittance as a function of the position inside the electron beam. (Second Row, Left) Energy profile along the beam. (Second Row, Right) Electron beam energy spread profile. (Bottom row) Resistive wakefields in the SASE3 undulator.}
\label{fig:ebeam}
\end{figure}

\begin{figure}
	\centering
	\includegraphics[angle=90,width=1.1\textwidth]{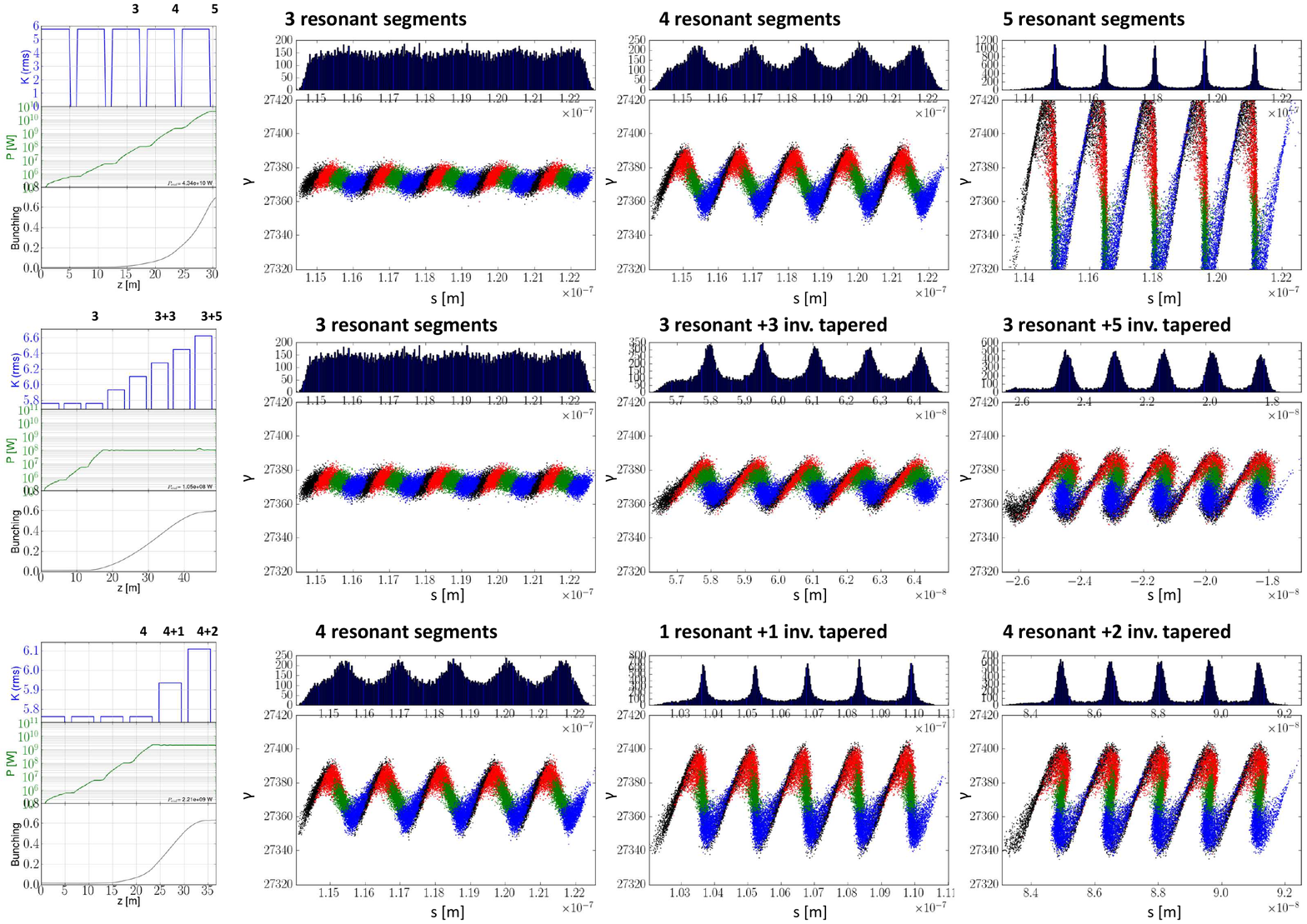}
	\caption{Electron phase space at various positions inside the SASE3 undulator resonant at 800~eV photon energy.}
	\label{fig:el_ph_sp}
\end{figure}

As discussed above, the inverse tapering technique allows one to obtain a high degree of electron density modulation in the beam, while significantly reducing the FEL radiation power. In this study we used the electron beam obtained from start-to-end simulations, shown in Figure~\ref{fig:ebeam}. We found that the optimal inverse tapering strategy for our electron beam parameters and the SASE3 undulator at the European XFEL ($ \lambda_w=68 $~mm, 21 segments, 5~meter-long each with 1.1~meter intersections) is to start the SASE FEL process with a uniform undulator up to the point when the bunching reaches the value of $0.025-0.05$ (radiation power is 3-4 orders of magnitude below saturation). Then a linear inverse tapering law is introduced, such that the undulator K value is increased by about 3\% per undulator section. In this case the radiation power growth is suppressed, while bunching grows linearly.
In Figure~\ref{fig:el_ph_sp} we show the evolution of the electron phase space for different undulator configurations. These configurations lead to a comparable bunching at the fundamental harmonic, but qualitatively different electron phase space distributions in the bucket at the undulator end. Time-dependent simulations were run for the nominal 20~pC electron beam with an energy of 14~GeV. The beam was propagated trough the SASE3 undulator segments resonant at 800~eV photon energy.
When no inverse tapering is applied (Fig.~\ref{fig:el_ph_sp}, top row), the bunching value of~0.6 is reached within 5 undulator segments, generating 40~GW FEL power. If inverse tapering is introduced after the $3^{rd}$ undulator (Fig.~\ref{fig:el_ph_sp}, middle row), a total number of 8 undulators is required to reach the same bunching value with 100~MW radiation power. Finally, introducing inverse tapering after the $4^{th}$ undulator (Fig.~\ref{fig:el_ph_sp}, bottom row), only a total of 6 undulator segments is needed to reach maximum bunching at the expense of a higher FEL output (2~GW) and a larger electron beam energy spread.

We now proceed investigating the performance of the baseline SASE3 beamline of the European XFEL. Four 2~m-long segments of a helical undulator with $ \lambda_w=80 $~mm period are assumed to be installed at the end of the planar baseline undulator\footnote{We assume that the helical undulator segments are installed in the space of the two last baseline segments which in turn are reinstalled at the beginning of SASE3, (see Figure~\ref{fig:schemes}-b)}. As a simplifying assumption, we consider the radiation from a helical undulator to be perfectly circularly polarized.

For our case study we set the electron beam to 12~GeV energy for simulating photon energies of 500~eV, 1~keV and 2~keV, and the 8.5~GeV for 250~eV photon energy.

\begin{figure}
	\centering
	\includegraphics[width=0.49\textwidth]{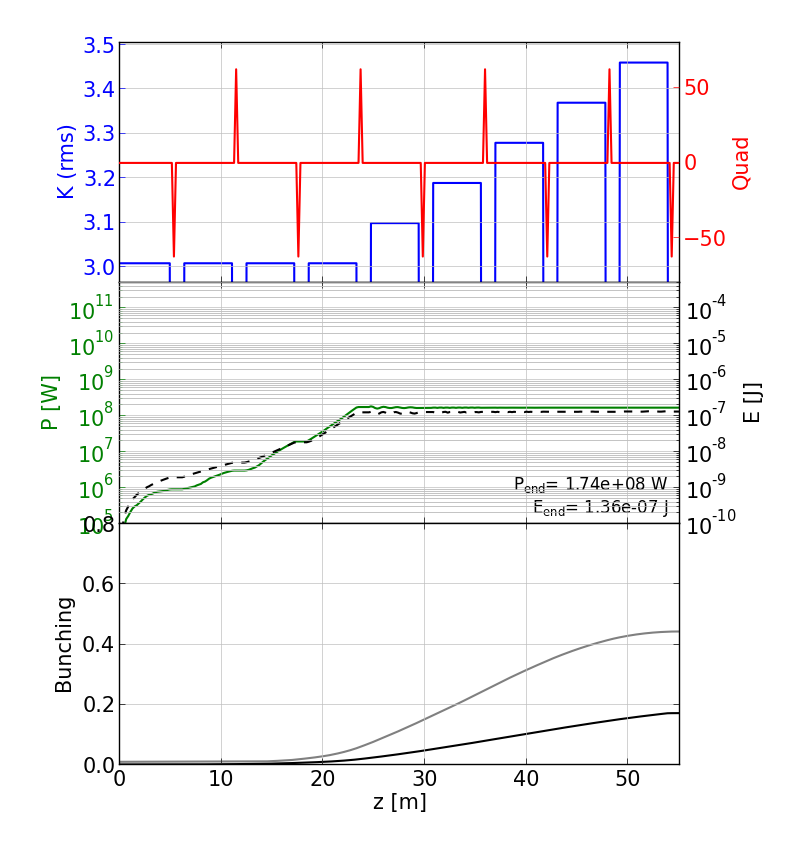}
	\includegraphics[width=0.49\textwidth]{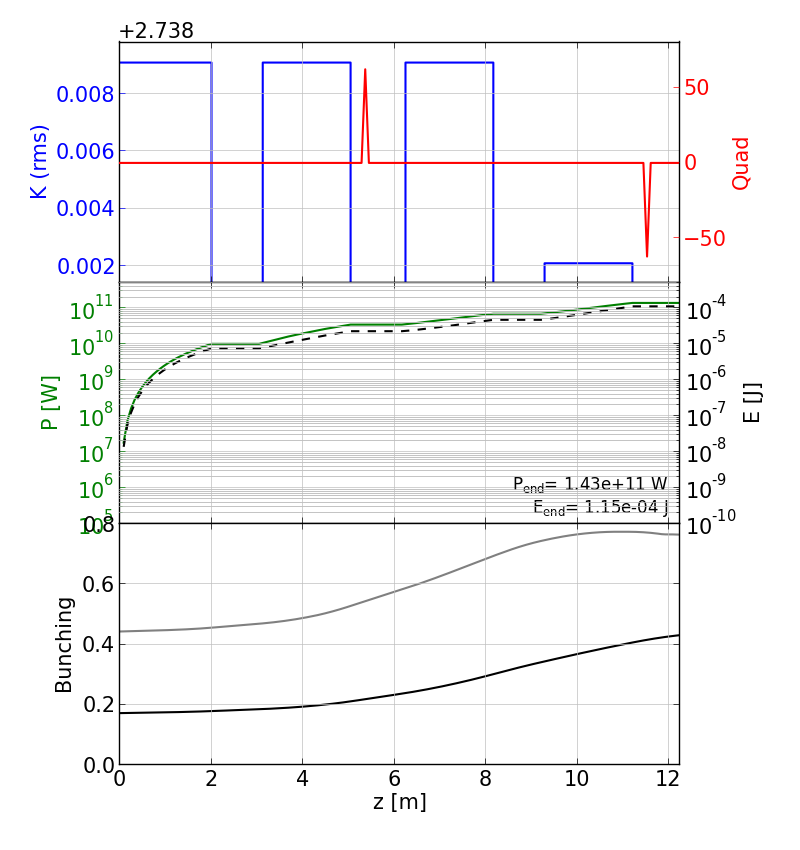}
	\caption{Electron beam and radiation parameters evolution within planar (left subfigure) and helical (right subfigure) undulators for 2000~eV photon energy. Top subplots represent undulator K values and quadrupole fields. Next subplots provide the radiation peak power and integrated energy of the pulse as a function of the undulator length. Last subplots show the evolution of a maximum (black line) and average (grey line) bunching along the beam}
	\label{fig:2000eV_evo}
\end{figure}

Electron beam and radiation evolution inside the inverse tapered planar undulator are presented in the left plot of Figure~\ref{fig:2000eV_evo} for the 2000~eV photon energy case. 
The electron beam distribution with developed microbunching is dumped and used as an input for the next simulation stage -- radiating in the helical undulator. We found that a free-space drift of the electron beam between the inverse tapered undulator and the helical one, does not significantly affect the electron beam properties, and is therefore ignored in our numerical simulations. 
In the helical undulator the already-bunched electron beam quickly reaches saturation, therefore the appropriate linear post-saturation tapering is applied. Simulation results for the helical radiator are presented in the right plot of Figure~\ref{fig:2000eV_evo}.

Numerical studies indicate that by means of inverse tapering of the baseline undulator one can substantially decrease the linearly polarized radiation output power. The energy of the 2~keV radiation pulse produced in the helical undulator (0.11~mJ) outruns the linearly polarized one (0.1~$\mu$J) by more than 3 orders of magnitude (see Table~\ref{table:results}, last column). 
A high degree of circular polarization of the total radiation pulse is thus obtained. 

Results for the other energy points are summarized in Table~\ref{table:results} (see inverse tapered contribution). It is important to remark, that at low photon energies there are several FEL gain lengths within a single undulator segment and it becomes challenging to stop the FEL amplification at the certain desired power level, should it be reached in the middle of an undulator. When we calculated our results we did not account for this effect. However, in order to reach better performance one may detune the first undulator segment.
It is also worth mentioning that an increase of the output linearly polarized radiation divergence takes place when the number of uniform undulator segments is reduced.

\section{Complementary methods to increase the degree of circular polarization}

\begin{figure}
	\centering
	\includegraphics[width=1.0\textwidth]{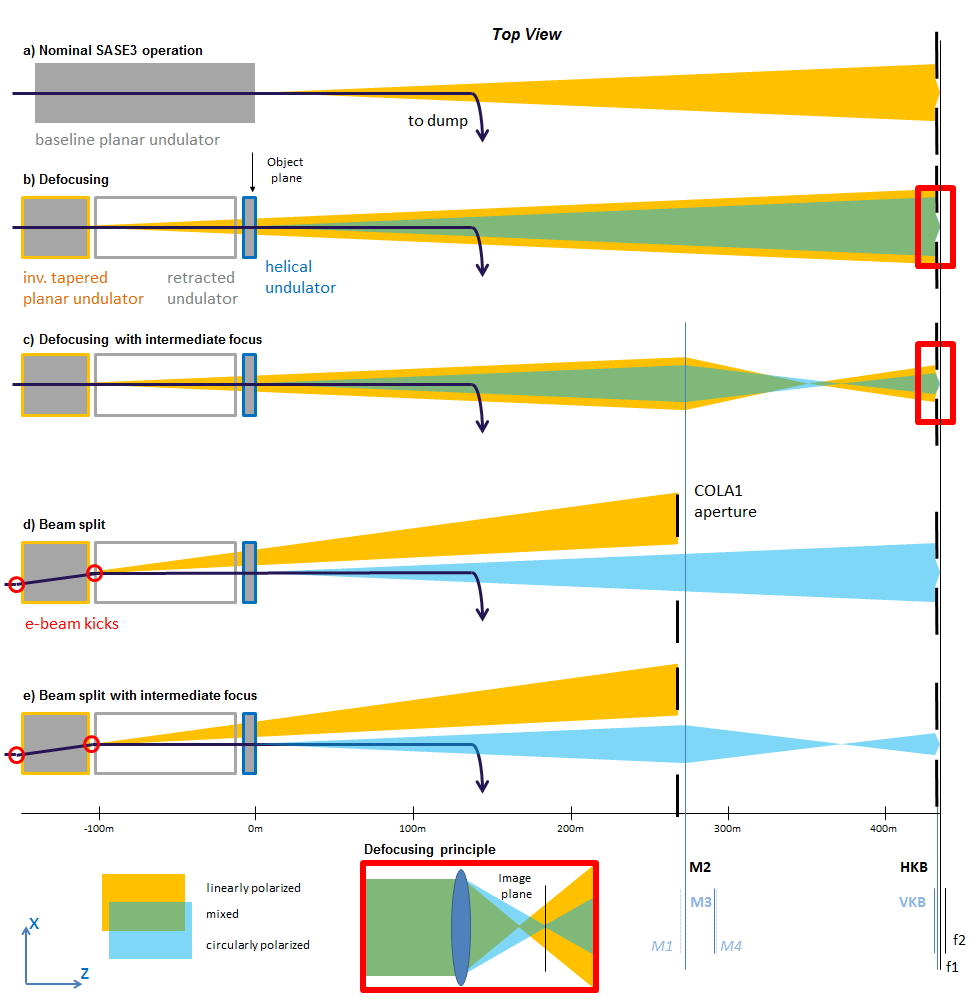}
	\caption{Schematic illustration of methods to generate and deliver circularly polarized radiation to the sample.}
	\label{fig:schemes}
\end{figure}

As discussed above, the inverse tapering technique is expected to be very efficient at the SASE3 line of European XFEL. In this section we discuss complementary but independent methods, to disentangle the residual linearly polarized component in the output radiation pulse. These methods can be used  in combination with inverse tapering to guarantee the stable delivery of a high-degree circular polarization to the SASE3 scientific instruments. We also investigate a way to optimize ratio of the circularly polarized component of the photon density at the sample over the linearly polarized one. 
Simulations in this section are based on the wavefront propagation technique, implemented in the code Synchrotron Radiation Workshop~\cite{Chubar1999}. The FEL radiation sources are modeled as Gaussian sources, based on the FEL radiation divergence. In order to disentangle the effects of our methods from those of inverse tapering, we model the sources of linearly and circularly polarized pulses with the same intensity, therefore assuming no linear polarization suppression via inverse tapering.
The radiation is propagated through the optics beamline layout down to the interaction region~\textit{f1} of the SQS instrument where the sample would be introduced (see Figure~\ref{fig:schemes}). No imperfections of the optical components were assumed during simulations in order to study solely the effects of linearly polarized background suppression methods.

Only the focusing optical components were modeled for radiation propagation, such as the offset mirror M1, the monochromator pre-mirrors M3a and M3b and the SQS KB mirror pair. The KB mirrors of the SQS instrument are tuned such that the radiation originated in the helical undulator is focused on the sample.
Locations and parameters of the optical components that we use for the simulations are provided in Table~\ref{table:optics}.

We assume that no circular birefringence effects take place after radiation reflection from the beamline components.

\begin{table}
\makegapedcells
\setcellgapes{3pt}
\scriptsize
\centering
	\begin{tabular}{cc}
		\hline\hline
		                Parameter        &        Value      \\ \hline
		 \multicolumn{2}{c}{\textbf{M1 - plane mirror (ignored)}}\\ \hline
		 \multicolumn{2}{c}{\textbf{M2 - bendable mirror}}\\ \hline
	focusing plane       &    horizontal      \\ \hline
	position$^1$      &     272~m      \\ \hline
	length        &       850~mm      \\ \hline
	radius of curvature        &      \makecell{8200~m (LE$^2$)\\ 27333~m (HE$^3$)}       \\ \hline
	inc. angle        &      \makecell{20~mrad (LE)\\ 6~mrad (HE)}       \\ \hline
		 \multicolumn{2}{c}{\textbf{M3 - cyllindrical mirror}}\\ \hline
	focusing plane       &    vertical      \\ \hline
	position       &     288~m      \\ \hline
	length        &       580~mm      \\ \hline
	radius of curvature        &      \makecell{7482~m (LE)\\ 16710~m (HE)}       \\ \hline
	inc. angle        &      \makecell{20~mrad (LE)\\ 9~mrad (HE)}       \\ \hline

	\end{tabular} \hspace{1cm}
	\begin{tabular}{cc}
\hline
		 \multicolumn{2}{c}{\textbf{M4 -plane mirror (ignored)}}\\ \hline
		 \multicolumn{2}{c}{\textbf{Exit slit (opened)}}\\ \hline
	position       &    388~m      \\ \hline
		 \multicolumn{2}{c}{\textbf{VKB - elliptical adaptive mirror}}\\ \hline
	focusing plane       &    vertical      \\ \hline
	position       &    430.6~m      \\ \hline
	length        &       800~mm      \\ \hline
	inc. angle        &      9~mrad       \\ \hline
		 \multicolumn{2}{c}{\textbf{HKB - elliptical adaptive mirror}}\\ \hline
	focusing plane       &    horizontal      \\ \hline
	position       &   431.4~m      \\ \hline
	length        &      800~mm      \\ \hline
	inc. angle        &     9~mrad       \\ \hline
		\multicolumn{2}{c}{\textbf{f1 - image plane}}\\ \hline
      	position       &   433.2~m      \\ \hline  
		\multicolumn{2}{c}{\textbf{f2 - image plane}}\\ \hline
      	position       &   435.23~m      \\ \hline\hline

	\end{tabular}
	\caption{Optical system parameters used for the calculations, based on \cite{Mazza2014a}
\newline $^1$ distance from the last undulator segment end \newline $^2$ low energy operation geometry \newline $^3$ high energy operation geometry}
	\label{table:optics}
\end{table}

\subsection{Defocusing approach}

In order to carry out certain experiments, it may be enough to \textit{reduce the photon density} of the linearly polarized radiation on the sample below a certain threshold, obtaining an acceptable degree of circular polarization. 
Since the planar and helical undulators may be separated spatially by a distance much larger than the Rayleigh length of the emitted radiation, one can obtain two separate sources of linearly and circularly polarized radiation. 
We separately studied two cases when an intermediate focus (IMF) is present, or not (see Fig.~\ref{fig:schemes} - b and c).

\begin{figure}
	\scriptsize
	\centering
	\hspace{0.8cm} circular at sample \hspace{1.4cm} linear at sample \hspace{1.3cm} linear at its waist \\
	\rotatebox{90}{\hspace{1cm}without IMF}
	\includegraphics[width=0.25\textwidth]{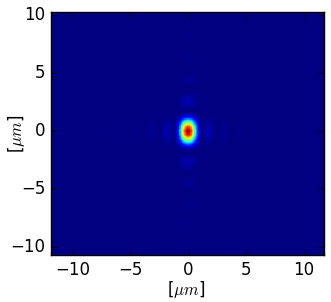}
	\includegraphics[width=0.25\textwidth]{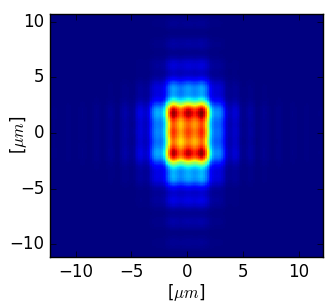}
	\includegraphics[width=0.25\textwidth]{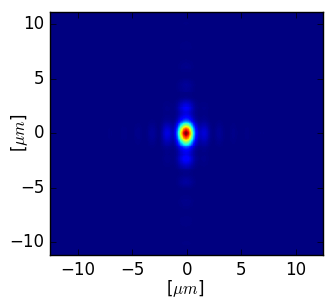}

	\rotatebox{90}{\hspace{1cm}with IMF}
	\includegraphics[width=0.25\textwidth]{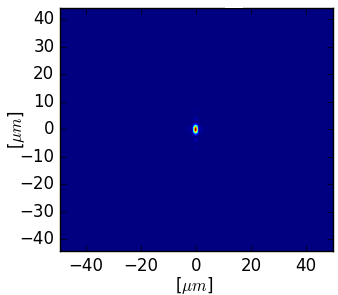}
	\includegraphics[width=0.25\textwidth]{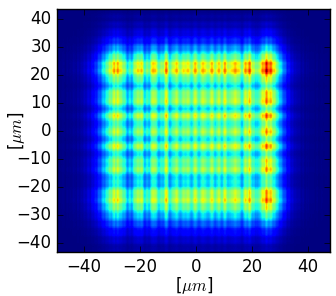}
	\includegraphics[width=0.25\textwidth]{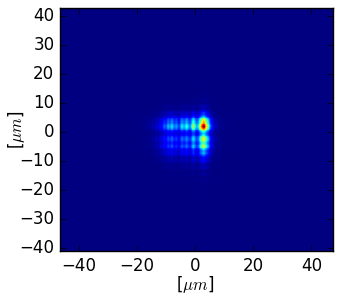}
	
	\caption{
Transverse intensity distributions of the 250~eV radiation, obtained without (first row) and with (second row) introduction of an intermediate focus (IMF) with transverse sizes of $1.1\times1.5~\mu m^{2}$ and $1.3\times2.3~\mu m^{2}$ correspondingly (full width at half maximum). First column - circularly polarized radiation distribution at the sample position, second column - out-of-focus linearly polarized radiation distribution at the same position. Third column - the linearly polarized radiation distribution at its minimum waist, located 2~mm upstream of the sample if no IMF introduced, and 22~mm upstream in the other case. \newline
The photon density ratio between the circularly and linearly polarized radiation at the sample is  approximately $5\cdot10^{-2}$ without IMF and $1\cdot10^{-3}$
Mirrors height error effects are ignored.
}
	\label{fig:Defocused}
\end{figure}

Let us first consider the case when no IMF is present (Fig.~\ref{fig:schemes}-b). 
The two sources will inevitably be focused by the SQS KB system to two separated images, located nearly 2~mm apart. At the position when one of the images is focused (in our case - the circularly polarized radiation), the other would be out of focus, forming a plateau with significantly lower photon density, as presented in Figure~\ref{fig:Defocused} (first row) for the case of 250~eV. 
Hereinafter we refer to this method of reducing the photon density of linearly polarized background due to defocusing as the \textit{defocusing approach}, since the linearly polarized radiation is out-of-focus at the sample plane. 
The resulting photon density ratio of the two polarization components is $ 4.9\cdot10^{-2} $ for 250~eV (see Table~\ref{table:results}, photon density ratio, no kick, no IMF). At the higher energy of 2~keV instead, the same ratio amounts to $ 1.7\cdot10^{-3} $.

From these numbers we see that the efficiency of the defocusing approach changes with the photon energy. 
In fact, the geometrical transmission of the SASE3-SQS beamline varies: it reaches 80\% above 2~keV photon energy, but deteriorates down to 5\% at 250~eV, since KB focusing mirrors become overfilled with highly divergent radiation. These values are slightly lower than the nominal ones, presented in~\cite{Mazza2012} (Fig.~3.2.3, red curve) due to larger-than-nominal divergence of the inverse-tapered FEL radiation. 
The finite size of the projected clear aperture of the KB mirrors increases the Rayleigh length of the focused radiation waist or, in other words, the depth of focus. When the Rayleigh length of the waists becomes comparable or smaller than their longitudinal separation, the defocusing approach is not effective any more. 
Fortunately, a beam transport scheme with an IMF is foreseen via insertion of mirrors M3-M4 to the beam path and changing the curvature of the initially plane mirror M2. While mirrors M1 and M4 remain flat and only direct the FEL radiation, mirrors M2 and M3 focus the radiation in the horizontal and vertical planes, creating the IMF (see Fig.~\ref{fig:schemes}-c).

The IMF introduction allows
\begin{itemize}
	\item to transport long wavelength radiation through the KB mirrors much more efficiently;
	\item to increase the distance between images of circularly and linearly polarized radiation.
\end{itemize}

\begin{figure}
	\centering
	\includegraphics[width=0.6\textwidth]{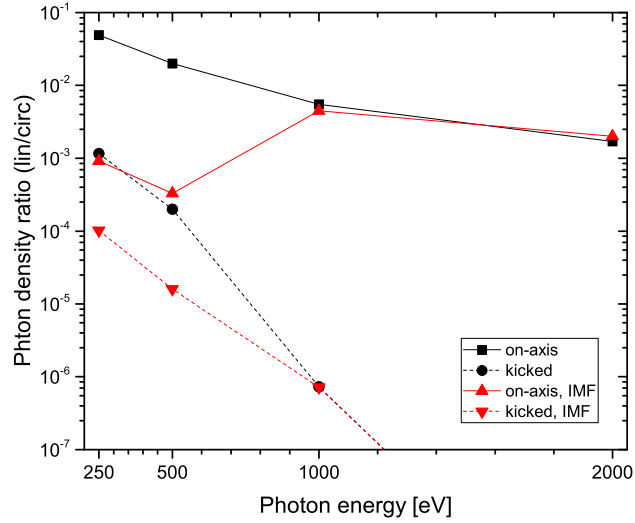}
	\caption{Reduction of the peak photon density at the sample due to defocusing effect for different radiation propagation scenarios:
	 ``on-axis'' stands for the nominal SASE3 radiation propagation scenario without neither e-beam kick, nor IMF introduction;
	 ``IMF'' indicates the presence of the intermediate focus in the radiation propagation geometry;
	 the electron beam, ``kicked'' transversely in the undulator by the maximum possible angle radiates in the propagation direction, reducing the geometrical beamline transmission. No inverse tapering contribution is assumed: sources of equal intensities are modeled.
	}
	\label{fig:ph_dens}
\end{figure}

The introduction of an IMF increases the distance between waists of different polarizations in terms of their Rayleigh lengths. It results in a larger spot size difference, as presented in Figure~\ref{fig:Defocused}, bottom row. Therefore, IMF allows one to increase the ratio of the photon density of different polarizations by more than one order of magnitude: from $ 4.9\cdot10^{-2} $ to $ 9.2\cdot10^{-4} $ at 250~eV (see Table~\ref{table:results}, photon density ratio, no kick, IMF). Photon density ratio is also presented graphically on Figure~\ref{fig:ph_dens} for different scenarios.
At higher photon energies the benefit of the IMF is not as strong: radiation divergence is small and beamline geometrical transmission is large even without it.
These findings, obtained for the~\textit{f1} image plane, also apply for the image plane~\textit{f2}.

\subsection{Beam split approach}

The defocusing method is based on the exploitation of existing components and we expect it can be routinely implemented during European XFEL operations. However, for some applications it may be important to prevent the linearly polarized background from entering the experimental area or to reduce its pulse energy at the sample position. This implies some kind of spatial filtering. 
The fist aperture (``COLB-1'' element), that can be potentially used for the spatial filtering of the linearly polarized background is located 187~m downstream the helical undulator. At that position, both circular and linear polarized radiation pulses diverge to a comparable transverse size. Hence, the method proposed in~\cite{Geloni2011b} is not applicable anymore. 
However, if the circularly and linearly polarized radiation pulses are emitted at a certain angle with respect to each other, they may be separated at a far zone. To this end, the electron beam should be transversely deflected somewhere between the baseline and helical undulators. Then the linearly polarized background may be blocked with any arbitrary beamline aperture: in fact all apertures are located in the far zone of the radiation.

This possibility was investigated earlier in terms of designing a beam transport system, capable of transporting the electron beam through the bend to the next undulator while preserving the microbunching~\cite{Li2010}.
In the light of delta undulator commissioning at LCLS~\cite{Nuhn2014}, it was found that the transverse deflection of the electron beam does not effectively deteriorate the FEL power radiated in an undulator downstream. In the current approach we take advantage of this effect, theoretical explanation of which is proposed in~\cite{Geloni2015,Geloni2015a,Geloni2016,Geloni2016b}.

In order to effectively separate the two pulses one should introduce a transverse deflection of the electron beam larger than the FEL radiation divergence. We can define a criterium of effective spatial separation of the two beams by requiring a deflection angle larger than the sum of full width at half maxima of the two radiation beams divergences.
For example, the average 500~eV radiation divergence if 42~$\mu$rad FWHM, therefore in order to disentangle the two pulses spatially in the far zone, an 84~$\mu$rad transverse kick should be applied to the electrons (see Table~\ref{table:results}).

\begin{figure}
	\centering
	\includegraphics[trim= 50 300 730 60,clip,height=0.4\textwidth]{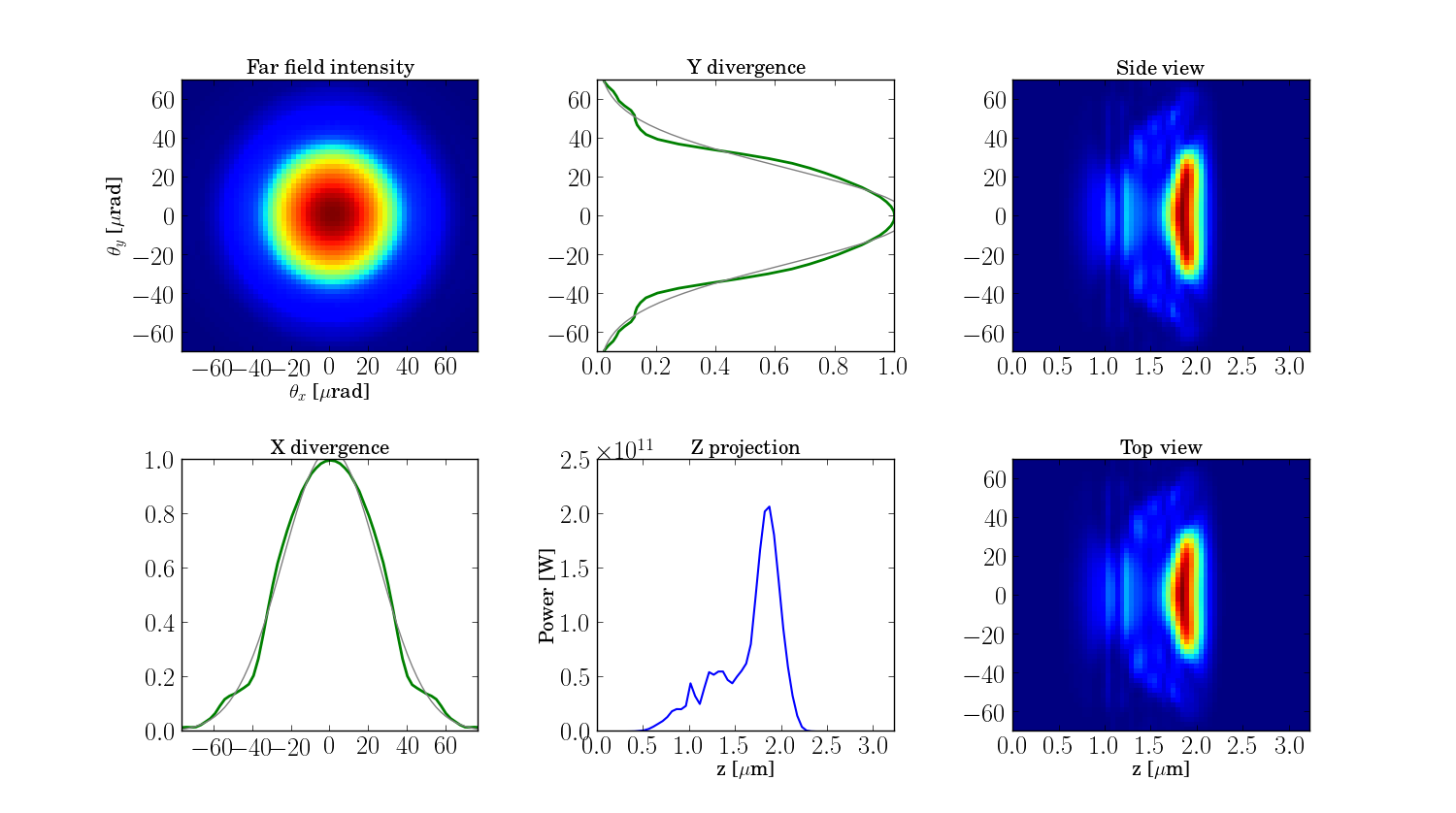}
	\includegraphics[trim= 100 300 730 60,clip,height=0.4\textwidth]{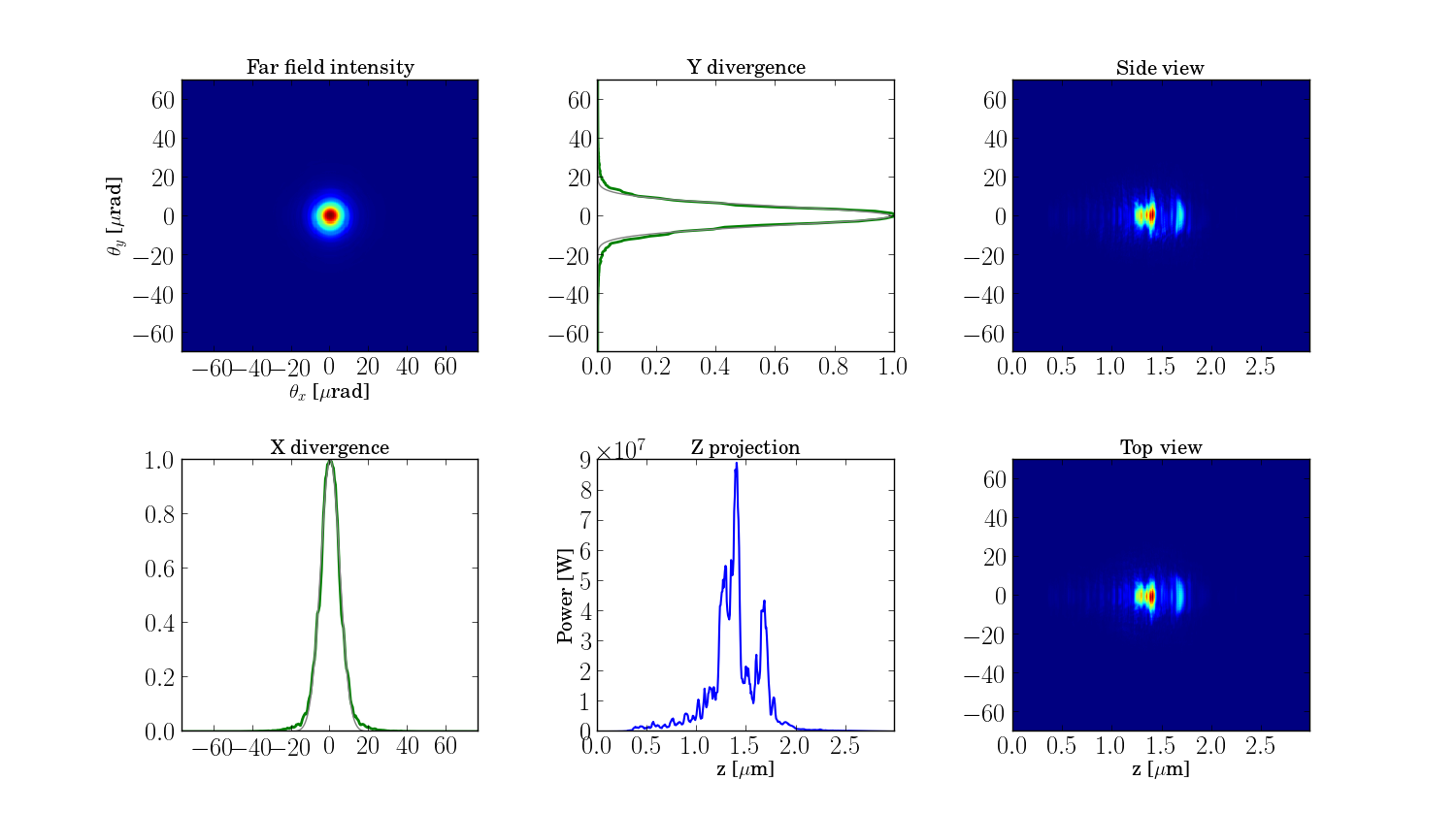}
	\caption{Angular intensity distribution of the far zone of the radiation from the inverse tapered undulator for 250 (left plot) and 2000~eV (right plot) photon energies. Simulation results provided for the ``typical'' shots.}
	\label{fig:rad_div}
\end{figure}

\begin{figure}
	\centering
	\includegraphics[width=0.6\textwidth]{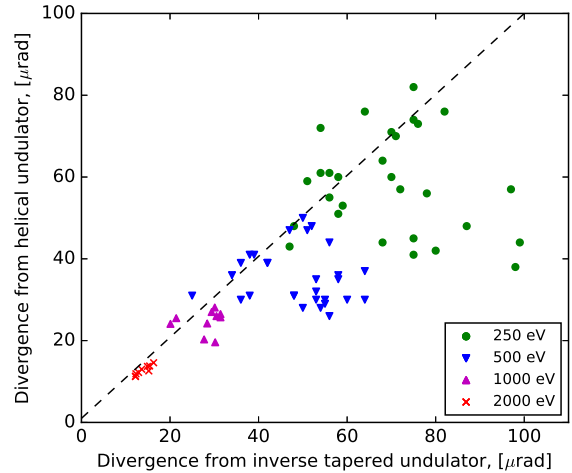}
	\caption{Divergence of the radiation, originating from the inverse tapered planar undulator (linearly polarized) and the helical undulator for various photon energies.}
	\label{fig:div_stat}
\end{figure}

Divergence of the SASE3 radiation pulses is presented in Figure~\ref{fig:rad_div} for two distant photon energies. In the SASE radiation mode the divergence may fluctuate significantly, and the linearly polarized radiation is more divergent on average (see Figure~\ref{fig:div_stat})

Once spatially separated, the linearly polarized radiation spot may be blocked with an aperture as discussed above. However, at low photon energies the required deflection of the electron beam becomes significant (see Table~\ref{table:results}).

The effective opening angle of the SASE3 beam transport aperture is comparable with the divergence of the FEL radiation at small photon energies (below 1~keV). Therefore, radiation propagating at an angle larger than 35~$\mu$rad with respect to the optical axis will be blocked by the 20~mm ``COLA1'' collimator located 3~m upstream the first offset mirror M1. The circularly polarized radiation of large wavelengths is then inevitably blocked if the orbit kick is applied within the helical undulator. Therefore, we limit ourselves to the scenario when the circularly polarized radiation is propagating parallel to the design optical axis and the transverse kick is applied to the electron beam within the inverse tapered undulator segments, which directs the linearly polarized radiation accordingly 
(see Figure~\ref{fig:schemes}-d).

\begin{figure}
	\centering
	\includegraphics[width=0.6\textwidth]{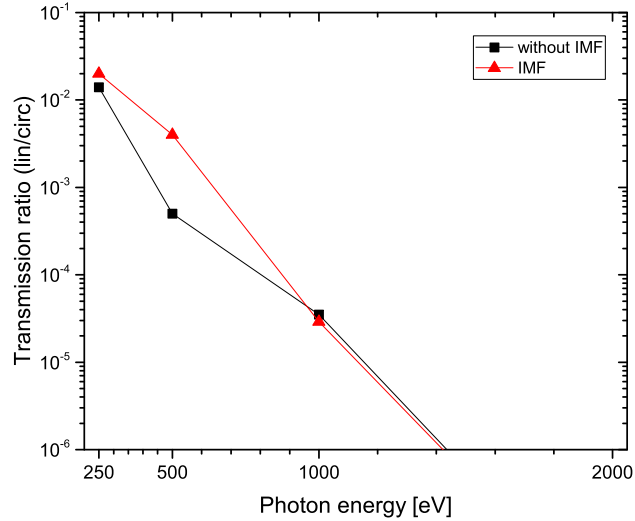}
	\caption{The ratio of beamline geometrical transmission of the linear over circular polarization for the maximum electron orbit kick inside the inverse tapered undulator. Equal source intensities are assumed.}
	\label{fig:Transm_ratio}
\end{figure}

The kick of the electron beam orbit implies an appropriate re-arrangement of the electron beam focusing system, namely - quadrupoles, in accordance to the new electron orbit. Quadrupoles should be shifted transversely according to a linear law to ensure that the electron beam passes through the optical center of each quadrupole. Due to the quadrupole displacement constraints (1.4~mm off the optical axis in either direction), we assume the maximum still-reasonable electron beam offset in both dimensions to be 2.5~mm (2$\times$1.25~mm).
Based on our numerical simulations, the required inverse tapered undulator length is e.g. 42.8~meters for the 500~eV photon energy case. This corresponds to a maximum possible deflection angle of 58~$\mu$rad, which is not enough for the effective linearly polarized radiation background elimination, while at 1~keV a maximum deflection of 26~$\mu$rad is just enough to fulfill our separation criterium (see Table~\ref{table:results}). Consequently the beam split approach is better used at the photon energies above 1~keV, where a maximum circular polarization can be reached (see Figure~\ref{fig:Transm_ratio}).

Finally at photon energies around 2~keV it may be already feasible to introduce a kick to the electron beam inside the helical undulator, accompanied with an appropriate transverse arrangement of the offset M1/M2 mirrors pair, which would significantly reduce the machine tuning time.


\begin{landscape}
\begin{table}
\makegapedcells
\setcellgapes{3pt}
\scriptsize 

\centering
	\begin{tabular}{ccccccccc}
		\hline\hline
Photon energy [eV] &       \multicolumn{2}{c}{ 250}        &        \multicolumn{2}{c}{500}      &        \multicolumn{2}{c}{1000}  &        \multicolumn{2}{c}{2000}        \\ \hline
Electron energy [GeV]&       \multicolumn{2}{c}{ 8.5}        &        \multicolumn{2}{c}{12}      &        \multicolumn{2}{c}{12}  &        \multicolumn{2}{c}{12}        \\ \hline
&      no IMF        &       IMF      &       no IMF        &       IMF  &   no IMF        &       IMF &   no IMF        &       IMF       \\ \hline
& \multicolumn{8}{c}{\textbf{Inverse tapering contribution}}\\ \hline
Inversely tapered planar undulators      &       \multicolumn{2}{c}{5 (30~m)}      &      \multicolumn{2}{c}{7 (42~m)}      &      \multicolumn{2}{c}{ 8 (48~m)}&      \multicolumn{2}{c}{ 9 (54~m)}      \\ \hline
Helical undultors      &      \multicolumn{2}{c}{ 4 (8~m)}      &     \multicolumn{2}{c}{ 4 (8~m)}    &      \multicolumn{2}{c}{ 4 (8~m) }  &      \multicolumn{2}{c}{ 4 (8~m) }     \\ \hline
Radiation power ratio lin./circ. [W]   &  \multicolumn{2}{c}{\makecell{$2.4\cdot10^{8}/1.7\cdot10^{11}$\\ ($1.6\cdot10^{-3}$)}}&  \multicolumn{2}{c}{\makecell{$8.7\cdot10^{7}/ 2.1\cdot10^{11}$ \\ ($4\cdot10^{-4}$)}} &  \multicolumn{2}{c}{ \makecell{$3.3\cdot10^{7}/1.4\cdot10^{11}$ \\ ($2.3\cdot10^{-4}$)}}  &  \multicolumn{2}{c}{ \makecell{$1.4\cdot10^{8}/1.1\cdot10^{11}$ \\ ($1.2\cdot10^{-3}$)}}  \\ \hline
Radiation pulse energy ratio lin/circ. [J]   &   \multicolumn{2}{c}{\makecell{$5.3\cdot10^{-7}/3.0\cdot10^{-4}$ \\ ($1.7\cdot10^{-3}$)}}&   \multicolumn{2}{c}{\makecell{$1.6\cdot10^{-7}/2.9\cdot10^{-4}$ \\ ($5.5\cdot10^{-4}$)}}&   \multicolumn{2}{c}{\makecell{$5.4\cdot10^{-8}/1.2\cdot10^{-4}$ \\ ($4.5\cdot10^{-4}$)}}  &   \multicolumn{2}{c}{\makecell{$1.0\cdot10^{-7}1.1/\cdot10^{-4}$ \\ ($9\cdot10^{-4}$)}} \\ \hline

& \multicolumn{8}{c}{\textbf{Optical beam transport geometrical transmission, no inverse tapering assumed$^1$}}\\ \hline
Radiation divergence, averaged [$\mu$rad]$^2$ &          \multicolumn{2}{c}{64}          &          \multicolumn{2}{c}{42}       &           \multicolumn{2}{c}{26}    &           \multicolumn{2}{c}{13}          \\ \hline
Required electron beam kick [$\mu$rad]&  \multicolumn{2}{c}{128} &  \multicolumn{2}{c}{84} & \multicolumn{2}{c}{52}& \multicolumn{2}{c}{26} \\ \hline
Maximum electron beam kick [$\mu$rad]&  \multicolumn{2}{c}{81}  &  \multicolumn{2}{c}{58}&  \multicolumn{2}{c}{52} &  \multicolumn{2}{c}{45} \\ \hline
Transmission, circ.			 	&  5\% &  38\%    & 12\%  &    68\%  &28\%  &    34\% &75\%  &    81\%     \\ \hline
Transmission, lin.			 	&  3\% &  11\%    & 8.3\%  &    42\%   & 20\%  &   21\%  & 62\%  &    61\%     \\ \hline
Transmission, lin., maximum kick			 	&  0.07\% &  1\%    & 0.06\%  &    0.3\%  & 0.001\%  &    0.001\%   & 0\%  &    0\%     \\ \hline


& \multicolumn{8}{c}{\textbf{Defocusing approach contribution, no inverse tapering assumed$^1$}}\\ \hline

Photon density ratio, without kick 	& $ 4.9\cdot10^{-2} $	&  $ 9.2\cdot10^{-4}$  &  $ 2\cdot10^{-2} $	& $ 3.3\cdot10^{-4}$   	& $5.5\cdot10^{-3}$ 	&4.5$\cdot10^{-3}  $  	& $1.7\cdot10^{-3}$ &$ 2\cdot10^{-3}  $   \\ \hline
Photon density ratio, maximum kick  	& $ 1.1\cdot10^{-3} $    	&  $ 1\cdot10^{-4}  $  &   $ 2\cdot10^{-4}$ 	& $ 1.6\cdot10^{-5} $	&  $7.3\cdot10^{-7}$ 	&  $7.2\cdot10^{-7}$ 	& 0 				& 0     \\ \hline

		& \multicolumn{8}{c}{\textbf{Total background reduction in combination with inverse tapering}}\\ \hline
Pulse energy ratio, maximum kick  &   $2.4\cdot10^{-5}$  & $ 4.5\cdot10^{-5}$ & $2.7\cdot10^{-6}$ & $2.4\cdot10^{-7}$ & $1.6\cdot10^{-8}$ & $1.3\cdot10^{-8}$ & 0 & 0     \\ \hline
Photon density ratio, without kick &  $8.3\cdot10^{-5}$  & $1.5\cdot10^{-6}$ & $1.1\cdot10^{-5}$ & $1.8\cdot10^{-7}$ & $2.5\cdot10^{-6}$ & $2\cdot10^{-6}$ & $1.5\cdot10^{-6}$ & $1.8\cdot10^{-6}$     \\ \hline
Photon density ratio, maximum kick &  $1.9\cdot10^{-6}$  & $1.7\cdot10^{-7}$ & $1.1\cdot10^{-7}$ & $8.8\cdot10^{-9}$ & $3.3\cdot10^{-10}$ & $3\cdot10^{-10}$ & 0 & 0     \\ \hline\hline

	\end{tabular}
	\caption{Summary of the obtained results for 4 photon energies. In both beam split and defocusing approaches the same energy per pulse for both polarizations is assumed.
\newline $^1$ Model Gaussian sources of equal intensities for both linearly and circularly polarized radiation. \newline $^2$ Average over 30 shots, intensity full width at half maximum}
	\label{table:results}
\end{table}
\end{landscape}

\section{Conclusions}

We confirm via numerical simulations that the inverse tapering technique~\cite{Schneidmiller2013,Schneidmiller2016} is expected to yield a high electron density modulation while significantly reducing the FEL radiation power at the SASE3 line of the European XFEL.
This way, the installation of a helical radiator would allow one to reach a high degree of circular polarization: the contribution of the linearly polarized background to the total FEL radiation intensity would be in order of~0.1\% along the design SASE3 photon energy range.

Several complementary methods can be used to further decrease of the linearly polarized background intensity.
The beam split approach allows one to spatially separate the background and collimate a large portion of it. It is the most effective above the 1~keV photon energy, where it allows one to obtain a degree of circular polarization, limited by the helical undulator properties.

Another way to effectively reduce the background contribution is to decrease its photon density on the sample with the defocusing approach. Being very simple in practice, it introduces a photon density ratio from 100 up to 2000 on top of the inverse tapering contribution.

We conclude that even in the event of lower-than-expected performance of the inverse tapering technique, synergy with the described methods will allow a sufficient contrast between linearly and circularly polarizations

\section{Acknowledgments}
We wish to thank Michael Meyer, Suren Karabekyan, Daniele La Civita, Nina Golubeva and Nikolay Smolyakov for useful discussions and Serguei Molodtsov for his interest in this work.

\bibliography{library}
\end{document}